\documentclass[showpacs,amsmath,superscriptaddress,reprint,aps]{revtex4-1}

\usepackage{graphicx}
\usepackage{hyperref}

\begin{document}

\title
{Organic molecular tuning of many-body interaction energies in air-suspended carbon nanotubes}
\author{S.~Tanaka}
\affiliation{Quantum Optoelectronics Research Team, RIKEN Center for Advanced Photonics, Saitama 351-0198, Japan}
\affiliation{Nanoscale Quantum Photonics Laboratory, RIKEN Cluster for Pioneering Research, Saitama 351-0198, Japan}
\author{K.~Otsuka}
\affiliation{Quantum Optoelectronics Research Team, RIKEN Center for Advanced Photonics, Saitama 351-0198, Japan}
\affiliation{Nanoscale Quantum Photonics Laboratory, RIKEN Cluster for Pioneering Research, Saitama 351-0198, Japan}
\author{K.~Kimura}
\affiliation{Surface and Interface Science Laboratory, RIKEN Cluster for Pioneering Research, Saitama 351-0198, Japan}
\affiliation{Department of Advanced Materials Science, School of Frontier Sciences, The University of Tokyo, Chiba 277-8561, Japan}
\author{A.~Ishii}
\affiliation{Quantum Optoelectronics Research Team, RIKEN Center for Advanced Photonics, Saitama 351-0198, Japan}
\affiliation{Nanoscale Quantum Photonics Laboratory, RIKEN Cluster for Pioneering Research, Saitama 351-0198, Japan}
\author{H.~Imada}
\affiliation{Surface and Interface Science Laboratory, RIKEN Cluster for Pioneering Research, Saitama 351-0198, Japan}
\author{Y.~Kim}
\affiliation{Surface and Interface Science Laboratory, RIKEN Cluster for Pioneering Research, Saitama 351-0198, Japan}
\author{Y.~K.~Kato}
\email[Corresponding author. ]{yuichiro.kato@riken.jp}
\affiliation{Quantum Optoelectronics Research Team, RIKEN Center for Advanced Photonics, Saitama 351-0198, Japan}
\affiliation{Nanoscale Quantum Photonics Laboratory, RIKEN Cluster for Pioneering Research, Saitama 351-0198, Japan}

\begin{abstract}
We investigate adsorption effects of copper phthalocyanine molecules on excitons and trions in air-suspended carbon nanotubes. Using photoluminescence excitation spectroscopy, we observe that exciton energy redshifts gradually with the molecular deposition thickness. The trion emission is also observed at large deposition amounts, which indicates charge transfer between the phthalocyanine molecules and carbon nanotubes. Analysis of the spectra for individual tubes reveal a correlation between the exciton-trion energy separation and the exciton emission energy, showing that the many-body interaction energies scale similarly with the molecular dielectric screening.   
\end{abstract}

\maketitle

Single-walled carbon nanotubes (CNTs) have attracted considerable attention due to their remarkable physical and electronic properties, and much effort has been devoted to functionalize CNTs for expanding their capabilities~\cite{Hirsch:2002, Miyauchi:2013,He:2017,Prato:2005}. In particular, non-covalent functionalization with organic molecules is a powerful strategy for developing CNT based devices such as photovoltaics and photodetectors~\cite{Hirsch:2009, Knorr:2011,Guldi:2010,Gruner:2006,Flavel:2017}. The interactions between the organic molecules and CNTs are considered to be less perturbative compared to covalent modification, and therefore the outstanding properties of CNTs can be preserved. Organic molecules such as porphyrin and phthalocyanine couple to nanotubes through $\pi$-$\pi$ interactions to modify the charge density of CNTs, while the emission energy is reduced due to the molecular screening~\cite{Deleporte:2010, Orellana:2012,Orellana:2014, Voisin:2011}. Furthermore, there are unique exciton dynamics at the organic molecule/CNT interface including energy and charge transfer~\cite{Voisin:2011,Bruce:2008,Thompson:2013,Prezhdo:2018,Nichola:2011}. 

In the measurements performed for nanotubes in solutions, however, the existence of solvent molecules inevitably complicate the interpretation by influencing the interactions. 
Additionally, the solvents themselves provide molecular screening, reducing the dielectric effects of the organic molecules. Investigation of molecular adsorption on air-suspended nanotubes would provide invaluable information towards fundamental understanding of the adsorption effects, as thermal evaporation techniques~\cite{Rosa:2008,Basiuk:2018} allow for molecular deposition without introducing other molecules. 
The use of pristine tubes should further enable drastic dielectric modification for developing non-covalently functionalized CNT devices.

Here, we demonstrate control over many-body interaction energies in air-suspended carbon nanotubes by copper phthalocyanine (CuPc) molecule adsorption. The molecules are deposited on chirality-assigned CNTs by thermal evaporation, and we perform photoluminescence (PL) spectroscopy of the CuPc/CNT hybrid. From the averaged PL and PL excitation (PLE) spectra for various chiralities, we find that the $E_{11}$ and $E_{22}$ resonances redshift with increasing deposition thickness. Furthermore, a new emission peak is observed at an energy below the $E_{11}$ peak, which is attributed to trion emission. Data from individual tubes reveal a good correlation between exciton-trion energy separation and the $E_{11}$ energy, suggesting the existence of a universal relationship.  We consider a model assuming power law scaling of the interaction energies with dielectric constant to explain the observed correlation.

\begin{figure}
\includegraphics{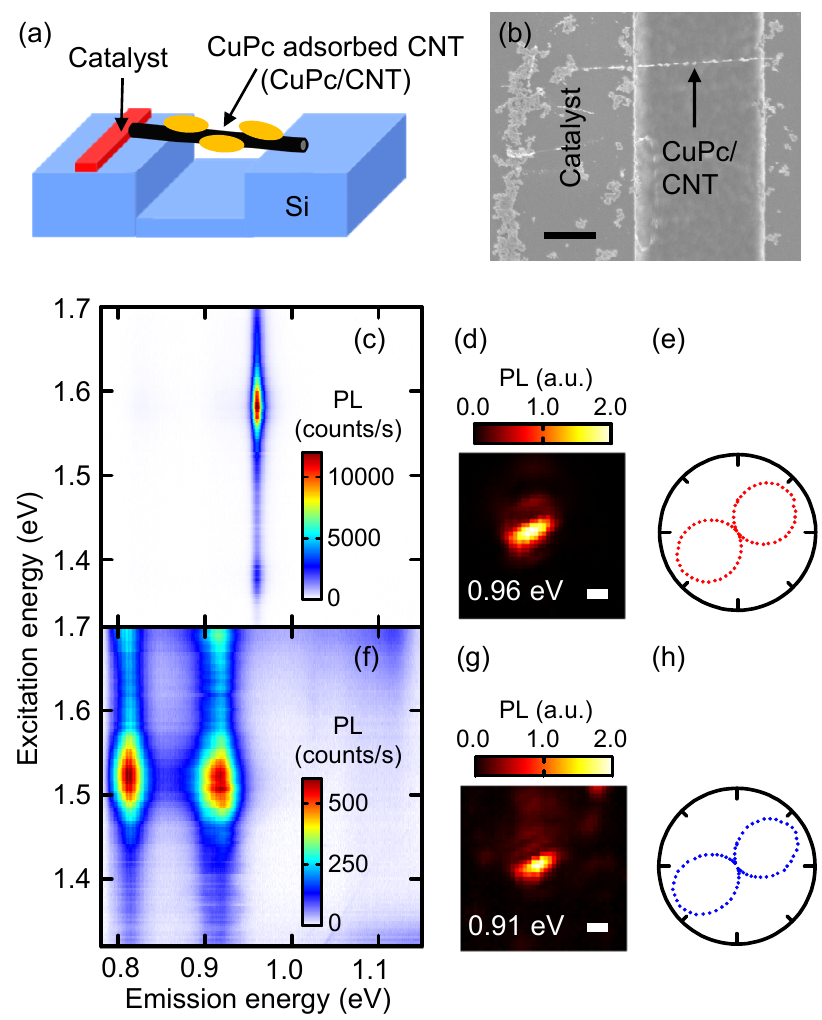}
\caption{
\label{Fig1} (a) A schematic image of our sample. (b) A scanning electron microscope image of a CuPc/CNT hybrid. (c), (d), and (e) PL excitation map, PL image, and laser polarization angle dependence, respectively, for  a (9,7) nanotube in the 26~nm sample before CuPc deposition. For (d) and (e), the PL intensity is obtained by integrating PL over a 10~meV wide spectral window centered at 0.96~eV. (f), (g), and (h) PL excitation map, PL image, and laser polarization angle dependence, respectively, for the same tube after the deposition. For (g) and (h), the PL intensity is obtained by integrating PL over a 10~meV wide spectral window centered at 0.91~eV. (b), (d), and (g) The scale bars are 1~$\mu$m.  (c)--(e) are taken with $P=5$~$\mu$W, and (f)--(h) are taken with $P=300$~$\mu$W.  
}
\end{figure}

Our suspended CNTs are grown over trenches on bare Si substrates [Fig.~\ref{Fig1}(a)]. We perform electron beam lithography and dry etching to form the trenches, and another electron beam lithography step defines catalyst areas near the trenches. Fe(III) acetylacetonate and fumed silica dispersed in ethanol are spin coated as catalyst for CNT growth. CNTs are synthesized by alcohol chemical vapor deposition at 800$^\circ$C for 1 minute~\cite{Ishii:2015}.

 We characterize the suspended nanotubes with a home-built sample scanning microspectroscopy system~\cite{Ishii:2015}. A wavelength tunable Ti:sapphire laser is used for excitation, and the laser polarization is rotated using a half-wave plate. The beam with a power $P$ is focused onto the nanotube by an objective lens, which is also used to collect the PL from the nanotubes. The PL signal is detected by an InGaAs photodiode array detector attached to a spectrometer. For PLE maps and PL images, the laser polarization is parallel to the tube axis, and all measurements are conducted at room temperature in dry nitrogen.

Carbon nanotubes are located by line scans along the trenches, and PLE measurements are performed. We check whether only a single peak is observed in the emission spectrum and whether the $E_{11}$ and $E_{22}$ resonance energies match tabulated data for air-suspended nanotubes~\cite{Ishii:2015}. If a nanotube satisfies these conditions, the position and the chirality for the nanotube is recorded into a list for further measurements. 

After the characterization, the sample is placed in a vacuum chamber for evaporation of CuPc (Sigma-Aldrich) and is kept at about 80$^\circ$C for ten minutes to remove air molecules before the evaporation. CuPc molecules are deposited on suspended CNTs in the chamber at room temperature using an evaporator heated to 480--520$^\circ$C under a vacuum of less than 10$^{-4} $~Pa. A glass slide is also placed in the chamber to quantify the deposition thickness from absorbance of CuPc peak at 2.0 eV~\cite{Tang:1986}. Calibration is performed by measuring the actual thickness by a surface profiler for two of the films. Four samples with different deposition amounts are prepared by changing the evaporation time, with nominal thickness on the substrate of 3 nm, 7 nm, 16 nm, and 26 nm. A scanning electron microscope image of a typical CNT after the evaporation is shown in Fig.~\ref{Fig1}(b). Bright spots are observed sparsely on the suspended CNTs, indicating inhomogeneous adsorption of CuPc. 

In order to investigate the molecular adsorption effects, we characterize the nanotubes after the deposition. We compare PL spectra, PL images, and polarization dependence for a suspended CNT before and after CuPc deposition for the 26 nm sample [Figs.~\ref{Fig1}(c)--(h)]. In the PLE map before the deposition [Fig.~\ref{Fig1}(c)], the emission occurs at the $E_{11}$ energy and the $E_{22}$ transition can be observed as a resonance in the excitation energy. The values for the $E_{11}$ and $E_{22}$ resonances are consistent with those for a (9,7) air-suspended nanotube~\cite{Ishii:2015}, indicating that air-molecules such as water are adsorbed~\cite{Uda:2018a,Lefebvre:2008}. Figure~\ref{Fig1}(f) shows a PLE map of the same tube after the deposition, and we find differences in the peak energy, the line width, and the intensity. Both of the $E_{11}$ and $E_{22}$ resonances show redshift as well as broadening, and the PL intensity decreases significantly with the molecular adsorption. Additionally, a new emission band appears in an energy region below the $E_{11}$ emission. The PL image of the molecular adsorbed nanotube is slightly smaller compared to that before the evaporation [Figs.~\ref{Fig1}(g),(d)], which is likely due to the inhomogeneous adsorption as observed in Fig.~\ref{Fig1}(b). In comparison, the laser polarization dependence does not change before and after the deposition [Figs.~\ref{Fig1}(e),(h)].

\begin{figure}
\includegraphics{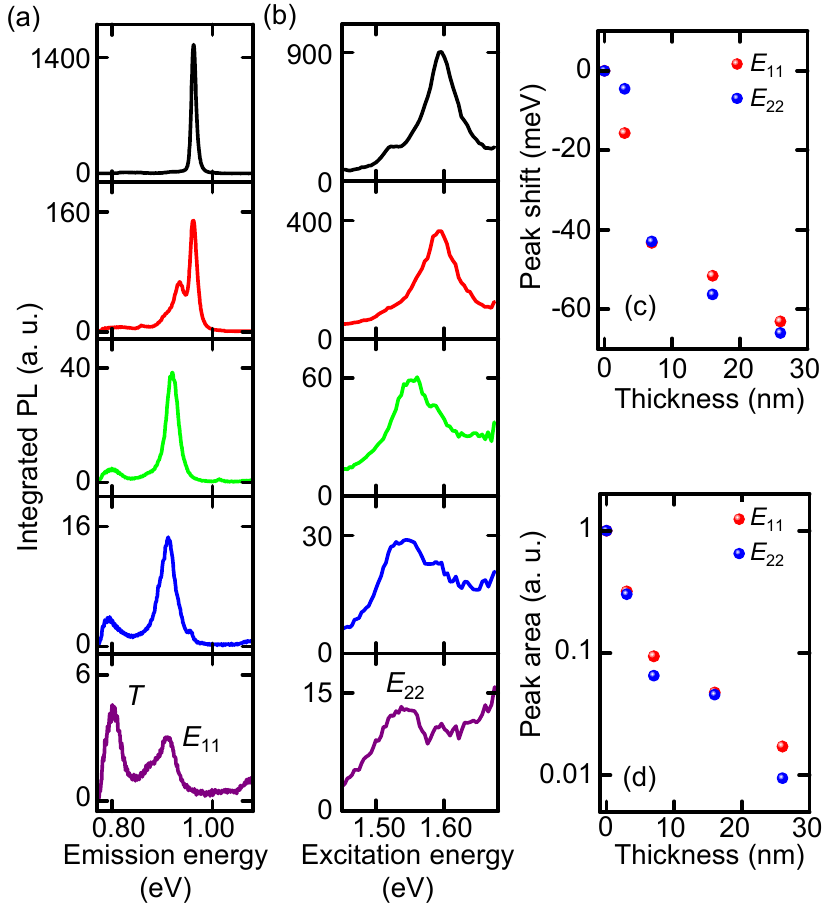}
\caption{
\label{Fig2} (a) and (b) Ensemble averaged PL and PLE spectra, respectively, for (9,7) tubes of the samples before deposition (black), 3~nm sample (red), 7~nm sample (green), 16~nm sample (blue), and 26~nm sample (purple). We measure 3--10 nanotubes for each sample, and check that the same tube is measured before and after the deposition from the location and the laser polarization dependence. The nanotubes before and after the deposition are measured with $P=100$~$\mu$W and $P=300$~$\mu$W, respectively. The peak shift and peak area obtained from the fitting are shown in (c) and (d), respectively. The peak shifts are compared to the peak energies for both resonances before the deposition, and the peak areas are normalized by those before the deposition.
}
\end{figure}

The PLE maps of the nanotubes for the four samples are measured to examine the deposition amount dependence of the adsorption effects. We integrate the PLE maps for each nanotube along the excitation and emission energy axes to obtain PL and PLE spectra, respectively, and the spectra for nanotubes with the same chiralities are averaged within each sample. The ensemble averaged PL and PLE spectra for (9,7) tubes are shown in Figs.~\ref{Fig2}(a) and~\ref{Fig2}(b), respectively. In the PL spectra [Fig.~\ref{Fig2}(a)], the intensity decreases and the $E_{11}$ resonance energy redshifts with the increase of the deposition thickness. Two peaks are observed for the 3~nm sample, corresponding to emission from regions of nanotubes with and without the CuPc molecule adsorption.
At large deposition amounts, the new emission band is observed as in Fig.~\ref{Fig1}(f). We will refer to this peak as the $T$ emission, whose intensity does not seem to depend much on the amount of CuPc molecules. In the PLE spectra [Fig.~\ref{Fig2}(b)], the $E_{22}$ resonance also redshifts with increasing deposition thickness, and the simultaneous redshifts for the $E_{11}$ and $E_{22}$ resonances can be explained by the molecular dielectric screening~\cite{Blackburn:2012,Lefebvre:2008}. For the 26~nm sample, the $E_{22}$ resonance peak is on top of a broad background slope, which may be attributed to the tail of the CuPc absorption peak at 2.0~eV~\cite{Tang:1986}. Exciton transfer from CuPc aggregates to CNTs would be consistent with the appearance of the absorption tail in the PLE spectra~\cite{Bruce:2008,Deleporte:2010}.

To evaluate the $E_{11}$ and $E_{22}$ resonance shifts for each sample, we extract the peak positions by fitting the ensemble averaged PL and PLE spectra. A Lorentz function is used to fit the $E_{11}$ resonance in the PL spectra, except for the 3~nm sample where two Lorentz functions are used to obtain a weighted average of the $E_{11}$ emission energy. In the case of the PLE spectra, we extract the $E_{22}$ resonance energy by fitting with a Lorentzian on top of a linear function. In Fig.~\ref{Fig2}(c), we plot the energy shifts of the $E_{11}$ and $E_{22}$ resonances for (9,7) nanotubes relative to the energies before the evaporation. They show similar behavior, and the energies decrease monotonically with the deposition thickness. The $E_{11}$ energy shifts by more than 60~meV, which is considerably larger than the shift caused by organic molecule adsorption in solutions ($\sim 20$~meV)~\cite{Deleporte:2010,Voisin:2011}.
The peak areas obtained from the fitting are also shown in Fig.~\ref{Fig2}(d), drastically decreasing as the thickness increases. A possible cause for the intensity reduction is carrier-induced PL quenching\cite{Yasukochi:2011,  Yoshida:2016, Uda:2018a} as charge transfer between a metal-phthalocyanine and a CNT has been predicted by a density functional calculation~\cite{Orellana:2012}.

\begin{figure}
\includegraphics{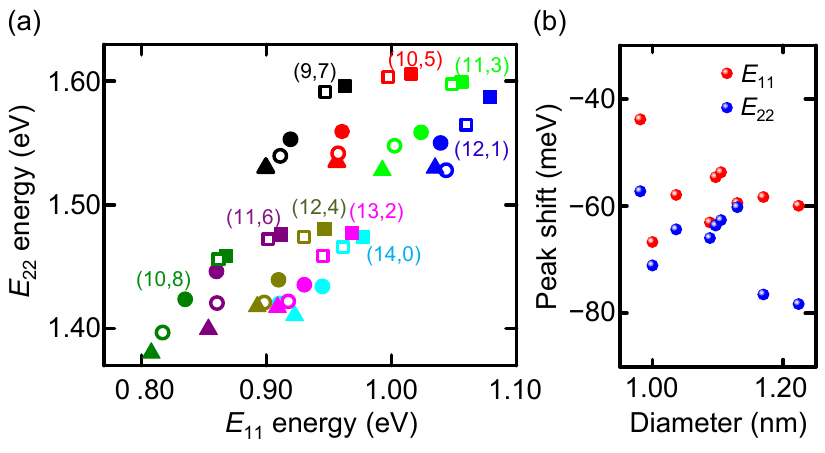}
\caption{
\label{Fig3} (a) Deposition thickness dependence of the $E_{11}$ and $E_{22}$ resonance energies for various chiralities. The data are obtained by the same procedure as in Fig.~\ref{Fig2}. The excitation energy integration windows are from 1.44~eV to 1.63~eV for (9,7), (10,5), (11,3), and (12,1) or from 1.35~eV to 1.50~eV for the others. If the $E_{11}$ resonance band is split into two as for the 3~nm sample in Fig.~\ref{Fig2}(a), we have obtained a weighted average of the emission energy by fitting the spectrum with two Lorentz functions. The integration window for the emission energy is from 0.77~eV to 1.09~eV for all chilarities. The filled squares, open squares, filled circles, open circles, and filled triangles correspond to the samples before deposition, 3~nm sample, 7~nm sample, 16~nm sample, and 26~nm sample, respectively. (b) The peak shifts of the $E_{11}$ and $E_{22}$ resonance energies for the 26~nm sample relative to those before the deposition are shown as a function of the diameter.     
}
\end{figure}

Using the same analysis procedure, the $E_{11}$ and $E_{22}$ resonance energies for various chiralities are obtained [Fig.~\ref{Fig3}(a)]. Redshifts of both resonances are observed for all chiralities measured, consistent with the dielectric screening effects due to the adsorbed molecules. The molecule-induced shifts of the $E_{11}$ and $E_{22}$ resonances for the 26~nm sample are plotted as a function of nanotube diameter in Fig.~\ref{Fig3}(b). Although we expect that the shifts become gradually smaller as the diameter increases~\cite{Nicholas:2009,Blackburn:2012,Strano:2007}, there is no clear dependence.  The scatter of the data points is likely caused by large tube-to-tube deviations of adsorption amounts. Averaging the shifts of the $E_{11}$ and $E_{22}$ resonances for all chiralities, we obtain values of $-58$~meV and $-67$~meV, respectively. Comparable emission energy shifts have been observed for nanotubes under dielectric screening by solvents~\cite{Ohno:2007,Blackburn:2012}, explaining why the shifts due to organic molecule adsorption in solutions are suppressed.

We evaluate the change in the exciton resonance energy by considering the reduction of the electronic many-body interactions due to molecular adsorption. The $E_{11}$ exciton energy is determined by $E_{11}=E_\text{sp}+E_\text{se}-E_\text{eb}$, where $E_\text{sp}$ is the single-particle bandgap, $E_\text{se}$ is the self-energy from repulsive electron interactions, and $E_\text{eb}$ is the exciton binding energy from the attractive electron-hole interactions~\cite{Swan:2007,Miyauchi:2015}. Because the magnitude of the self-energy is larger than that of the exciton binding energy, the $E_{11}$ exciton energy is higher than $E_\text{sp}$ by the net many-body correlation energy $E_\text{se}-E_\text{eb}$. Assuming that these interactions scale by the same factor $C_\text{X}$, the shift of the $E_{11}$ exciton energy $\Delta E_{11}$ relative to the air-molecule-adsorbed state is given by $\Delta E_{11} =(E_\text{se}-E
_\text{eb})\times (C_\text{X}-1)$. For air-molecule adsorbed nanotubes with 1-nm diameter, $E_\text{se}$ and $E_\text{eb}$ have been estimated to be 910~meV and 660~meV, respectively~\cite{Lefebvre:2008}. From the average $\Delta E_{11}$ of $-58$~meV for the 26~nm sample, $C_\text{X}$ is calculated to be 0.77, which corresponds to a reduction of the many-body interactions with the molecular adsorption by 23\%. The average $\Delta E_{22}$ is slightly larger than the average $\Delta E_{11}$, which is reasonable because the net many-body correlation energy for the $E_{22}$ excitons is larger than that for the $E_{11}$ excitons~\cite{Blackburn:2012, Dresselhaus:2007}. 

\begin{figure}
\includegraphics{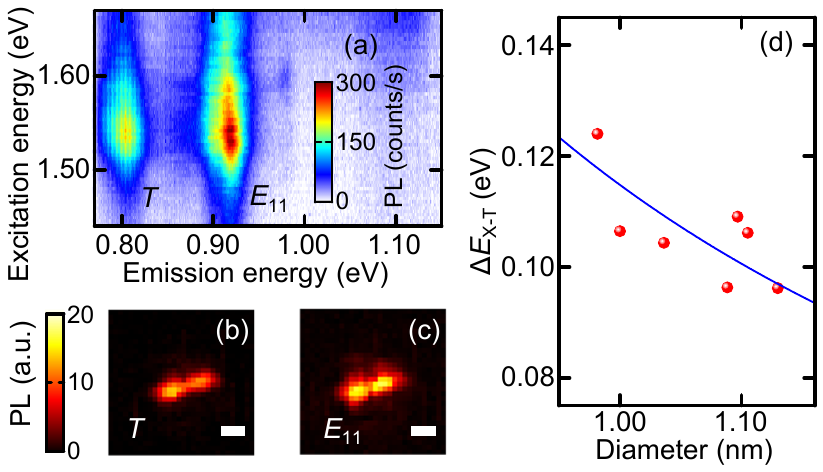}
\caption{
\label{Fig4} (a) A PLE map for a (9,7) tube in the 26~nm sample. (b) and (c) PL images for $T$ and $E_{11}$ exciton emission, respectively, for the same tube. 10~meV wide spectral integration windows centered at 0.81 and 0.92~eV are used for (b) and (c), respectively. (d) $\Delta E_\text{X-T}$ for the 26~nm sample is shown as a function of the diameter. All of the data are taken at $P=$300~$\mu$W.
}
\end{figure}

We now turn our attention to the low energy $T$ emission. Figure ~\ref{Fig4}(a) is a PLE map for a (9,7) nanotube in the 26~nm sample, and there are two peaks at 0.81~eV and 0.92~eV corresponding to the $T$ and $E_{11}$ emission, respectively. The absorption resonance energies for both emission are almost the same, and the PL images at the $T$ and $E_{11}$ energies overlap spatially [Figs.~\ref{Fig4}(b),(c)]. The spectral and spatial coincidence indicates that the $T$ emission comes from the same tube. 
We define the energy separation between the $T$ and $E_{11}$ peaks to be $\Delta E_\text{X-T}$, for example $\Delta E_\text{X-T}=0.11$~eV in Fig.~\ref{Fig4}(a). The $T$ emission energy for various chiralites are obtained by fitting the ensemble averaged PL spectra with a Lorentz function, and $\Delta E_\text{X-T}$ is plotted as a function of the diameter in Fig.~\ref{Fig4}(d). Since the $K$-momentum exciton emission should appear at 0.13--0.14~eV below the $E_{11}$ emission peak~\cite{Matsunaga:2010, Yoshida:2016}, it cannot explain the $T$ emission with $\Delta E_\text{X-T}=0.10$--0.12~eV. In addition, intensity of the $K$-momentum exciton emission is typically 1\% of the bright $E_{11}$ exciton emission~\cite{Yoshida:2016}, but the $T$ emission intensity can be comparable to the $E_{11}$ emission~\cite{Yoshida:2016}. A more plausible interpretation would be emission from trions, as the energy separation can be 0.10--0.14~eV under dielectric screening within the diameter range shown in Fig.~\ref{Fig4}(d)~\cite{Matsunaga:2011, Santos:2011}. The observed $\Delta E_\text{X-T}$ tends to become smaller as the diameter increases, consistent with the behavior for the exciton-trion energy separation~\cite{Yoshida:2016,Uda:2018a}. As a trion is a bound state of a carrier and an exciton~\cite{Matsunaga:2011}, we expect trion formation if there is charge transfer between CuPc molecules and CNTs. Such a picture can also explain the $E_{11}$ intensity reduction with the molecular adsorption.

The exciton-trion energy separation for the CuPc adsorbed state is smaller than that for the air molecule adsorbed state by about 70~meV for tubes with diameter $d=$1~nm, indicating that exciton-carrier interactions are considerably screened by the CuPc molecules. It is known that the exciton-trion energy separation $\Delta E_\text{X-T}$ follows the relation $\Delta E_\text{X-T}=E_\text{tb}+E_\text{st}$, where $E_\text{tb}=A/d$ is the trion binding energy and $E_\text{st}= B/d^{2}$ is the singlet-triplet splitting energy while $A$ and $B$ are proportionality constants~\cite{Matsunaga:2011,Yoshida:2016}. For the air molecule adsorbed state, $A=105$~meV$\cdot$~nm and $B=70$~meV$\cdot$~nm$^2$~\cite{Yoshida:2016}. We fit the data by $\Delta E_\text{X-T} = (A/d+ B/d^{2})\times C_\text{T}$, under an assumption that both $A$ and $B$ scale by the same factor $C_\text{T}$ upon the organic molecule adsorption. The result is shown as a blue line in Fig.~\ref{Fig4}(d), giving $C_\text{T} =0.67$, which is comparable to the value of $C_\text{X}$.

\begin{figure}
\includegraphics{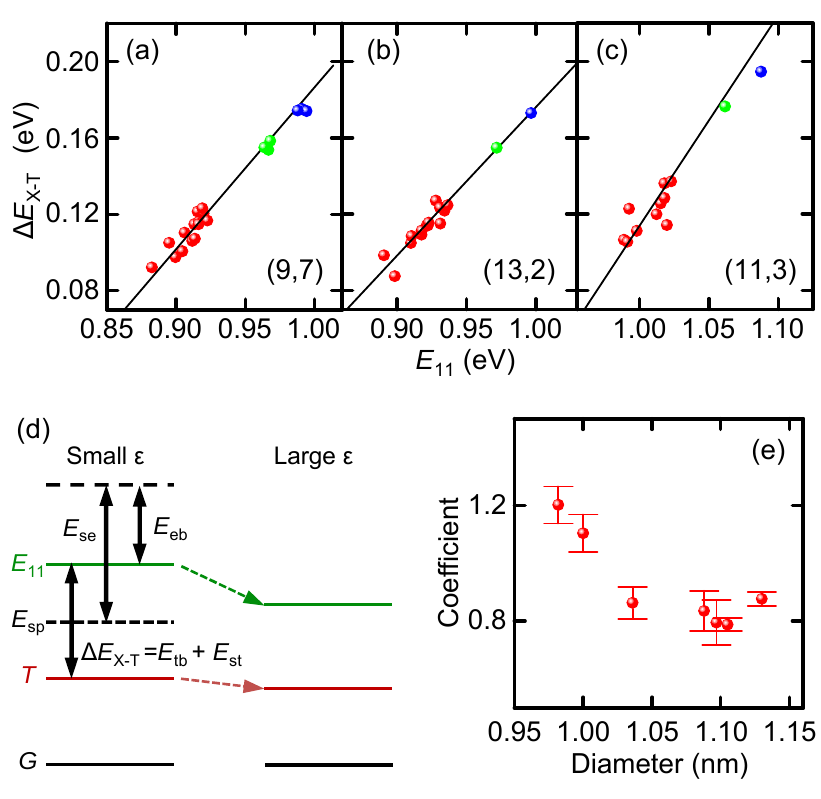}
\caption{
\label{Fig5} (a-c) Exciton-trion energy separation for (9,7), (13,2), and (11,3) tubes are shown as a function of the $E_{11}$ emission energy. Red, green, and blue points are the results for CuPc adsorbed states, air-molecule adsorbed states, and pristine states without molecular adsorption, respectively.  (d) Schematic diagram for tuning of the many-body interaction energies with the dielectric constant. (e) The coefficient for the correlation is shown as a function of the diameter.
}
\end{figure}

So far we have been discussing the ensemble spectra, but data for the individual tubes provide additional insights to molecular screening effects on the many-body interactions. By fitting the PL spectra for each nanotube, we obtain the energies for the $T$ and $E_{11}$ emission in various molecular screening environments. In Fig.~\ref{Fig5}(a), $\Delta E_\text{X-T}$ for (9,7) individual nanotubes are plotted as a function of the $E_{11}$ exciton emission energy. We find that there is a good correlation between $\Delta E_\text{X-T}$ and $E_{11}$, where $\Delta E_\text{X-T}$ becomes smaller as the $E_{11}$ emission energy redshifts. We also plot data points for (9,7) suspended nanotubes with and without air-molecule adsorption from ref~\citenum{Uda:2018a} as green and blue points, respectively. These data points also follow the trend, suggesting that there is a universal relation applicable to nanotubes under different surface conditions.  Similar correlation is observed for (13,2) and (11,3) nanotubes [Figs.~\ref{Fig5}(b) and~\ref{Fig5}(c), respectively], as well as for the other chiralities. We note that CuPc adsorption can shift the $E_{11}$ energy by $\sim 100$~meV compared to the pritine state, offering flexible tunability over the emission energy

The correlation can be understood in terms of the molecular dielectric screening [Fig.~\ref{Fig5}(d)]. In general, the many-body interactions determining the $T$ and $E_{11}$ exciton energies are sensitive to environment, which can simply be parameterized as the environmental dielectric constant $\varepsilon$. Assuming that interaction energies scale as $\varepsilon^{-\alpha}$ with $\alpha$ being a constant~\cite{Swan:2007,Lefebvre:2008}, $\Delta E_\text{X-T}=E_\text{tb}^{\varepsilon =1}\varepsilon^{-\alpha}+E_\text{st}^{\varepsilon =1}\varepsilon^{-\alpha}$ and $E_{11}=E_\text{sp}+E_\text{se}^{\varepsilon =1}\varepsilon^{-\alpha}-E_\text{eb}^{\varepsilon =1}\varepsilon^{-\alpha} $, where $E_\text{tb}^{\varepsilon =1}$, $E_\text{st}^{\varepsilon =1}$, $E_\text{se}^{\varepsilon =1}$, and $E_\text{eb}^{\varepsilon =1}$ are the trion binding energy, singlet-triplet splitting energy, self energy, and exciton binding energy at $\varepsilon =1$, respectively. The slope for the correlation is then given by $\frac{\partial \Delta E_\text{X-T}}{\partial E_{11}}=\frac{E_\text{tb}^{\varepsilon =1}+E_\text{st}^{\varepsilon =1}}{E_\text{se}^{\varepsilon =1}-E_\text{eb}^{\varepsilon =1}}$, indicating a linear relation between $\Delta E_\text{X-T}$ and $E_{11}$. The experimentally observed correlation therefore implies that the many-body interactions energies scale similarly within the experimental conditions explored, which is reasonable since the scaling factor for the $E_{11}$ exciton $C_\text{X}$ is comparable to that for the trion $C_\text{T}$.

Using the reported values for pristine suspended nanotubes of $E_\text{tb}^{\varepsilon =1}=117$~meV, $E_\text{st}^{\varepsilon =1}=78$~meV, $E_\text{se}^{\varepsilon =1}=980$~meV, and $E_\text{eb}^{\varepsilon =1}=725$~meV for 1~nm-diameter tubes~\cite{Lefebvre:2008,Uda:2018a}, the proportionality coefficient is about 0.8. To compare this value with the experimental results, we fit the data points for the organic molecule adsorbed state with a linear function. We impose the line to go through a data point for the air-molecule adsorbed state, where the $E_{11}$ emission energy is obtained from the table in ref~\citenum{Ishii:2015} and $\Delta E_\text{X-T}$ is determined from the equation given in ref~\citenum{Yoshida:2016}. The obtained coefficients range from 0.8 to 1.2 [Fig.~\ref{Fig5}(e)], which is a reasonable agreement considering the large uncertainty in the absolute value of $E_\text{se}^{\varepsilon =1}$. 

In summary, we have investigated adsorption effects of CuPc molecules on excitons and trions in air-suspended CNTs by measuring PL spectra of the CuPc/CNT hybrids. By averaging the PLE spectra for tubes with the same chiralities, the redshift of the $E_{11}$ and $E_{22}$ resonances have been observed with increasing deposition thickness due to the molecular screening. The $E_{11}$ energy modification from the pristine state to the CuPc adsorbed state can reach 100~meV, which is significantly larger than the energy shift with organic molecule adsorption in solutions. Furthermore, we find that the trion emission emerges at an energy lower than the $E_{11}$ exciton emission, which is likely due to charge transfer between CuPc molecules and CNTs. The spectra from individual tubes show that there is a good correlation between the exciton-trion energy separation and the exciton emission energy. A model assuming power law scaling of the many-body interactions with dielectric constant can quantitatively explain the observed correlation, which should be able to describe nanotubes with different surface conditions in a universal manner. Our findings show that organic molecule adsorption significantly affects many-body interaction energies of suspended CNTs, which opens up a pathway to CNT devices utilizing noncovalent molecular functionalization.

\begin{acknowledgments}
Work supported in part by JSPS (KAKENHI JP16H05962 and JP17H07359) and MEXT (Photon Frontier Network Program, Nanotechnology Platform).  We thank Advanced Manufacturing Support Team at RIKEN for technical assistance.
\end{acknowledgments}


\begin{thebibliography}{35}%
\makeatletter
\providecommand \@ifxundefined [1]{%
 \@ifx{#1\undefined}
}%
\providecommand \@ifnum [1]{%
 \ifnum #1\expandafter \@firstoftwo
 \else \expandafter \@secondoftwo
 \fi
}%
\providecommand \@ifx [1]{%
 \ifx #1\expandafter \@firstoftwo
 \else \expandafter \@secondoftwo
 \fi
}%
\providecommand \natexlab [1]{#1}%
\providecommand \enquote  [1]{``#1''}%
\providecommand \bibnamefont  [1]{#1}%
\providecommand \bibfnamefont [1]{#1}%
\providecommand \citenamefont [1]{#1}%
\providecommand \href@noop [0]{\@secondoftwo}%
\providecommand \href [0]{\begingroup \@sanitize@url \@href}%
\providecommand \@href[1]{\@@startlink{#1}\@@href}%
\providecommand \@@href[1]{\endgroup#1\@@endlink}%
\providecommand \@sanitize@url [0]{\catcode `\\12\catcode `\$12\catcode
  `\&12\catcode `\#12\catcode `\^12\catcode `\_12\catcode `\%12\relax}%
\providecommand \@@startlink[1]{}%
\providecommand \@@endlink[0]{}%
\providecommand \url  [0]{\begingroup\@sanitize@url \@url }%
\providecommand \@url [1]{\endgroup\@href {#1}{\urlprefix }}%
\providecommand \urlprefix  [0]{URL }%
\providecommand \Eprint [0]{\href }%
\providecommand \doibase [0]{http://dx.doi.org/}%
\providecommand \selectlanguage [0]{\@gobble}%
\providecommand \bibinfo  [0]{\@secondoftwo}%
\providecommand \bibfield  [0]{\@secondoftwo}%
\providecommand \translation [1]{[#1]}%
\providecommand \BibitemOpen [0]{}%
\providecommand \bibitemStop [0]{}%
\providecommand \bibitemNoStop [0]{.\EOS\space}%
\providecommand \EOS [0]{\spacefactor3000\relax}%
\providecommand \BibitemShut  [1]{\csname bibitem#1\endcsname}%
\let\auto@bib@innerbib\@empty
\bibitem [{\citenamefont {Hirsch}(2002)}]{Hirsch:2002}%
  \BibitemOpen
  \bibfield  {author} {\bibinfo {author} {\bibfnamefont {A.}~\bibnamefont
  {Hirsch}},\ }\bibfield  {title} {\bibinfo {title} {Functionalization of
  single-walled carbon nanotubes},\ }\href
  {https://www.onlinelibrary.wiley.com/doi/abs/10.1002/1521-3773%2820020603%2941%3A11%3C1853%3A%3AAID-ANIE1853%3E3.0.CO%3B2-N}
  {\bibfield  {journal} {\bibinfo  {journal} {Angew. Chem. Int. Ed.}\ }\textbf
  {\bibinfo {volume} {41}},\ \bibinfo {pages} {1853} (\bibinfo {year}
  {2002})}\BibitemShut {NoStop}%
\bibitem [{\citenamefont {Miyauchi}\ \emph {et~al.}(2013)\citenamefont
  {Miyauchi}, \citenamefont {Iwamura}, \citenamefont {Mouri}, \citenamefont
  {Kawazoe}, \citenamefont {Ohtsu},\ and\ \citenamefont
  {Matsuda}}]{Miyauchi:2013}%
  \BibitemOpen
  \bibfield  {author} {\bibinfo {author} {\bibfnamefont {Y.}~\bibnamefont
  {Miyauchi}}, \bibinfo {author} {\bibfnamefont {M.}~\bibnamefont {Iwamura}},
  \bibinfo {author} {\bibfnamefont {S.}~\bibnamefont {Mouri}}, \bibinfo
  {author} {\bibfnamefont {T.}~\bibnamefont {Kawazoe}}, \bibinfo {author}
  {\bibfnamefont {M.}~\bibnamefont {Ohtsu}}, \ and\ \bibinfo {author}
  {\bibfnamefont {K.}~\bibnamefont {Matsuda}},\ }\bibfield  {title} {\bibinfo
  {title} {Brightening of excitons in carbon nanotubes on dimensionality
  modification},\ }\href {\doibase 10.1038/nphoton.2013.179} {\bibfield
  {journal} {\bibinfo  {journal} {Nat. Photon.}\ }\textbf {\bibinfo {volume}
  {7}},\ \bibinfo {pages} {715} (\bibinfo {year} {2013})}\BibitemShut {NoStop}%
\bibitem [{\citenamefont {He}\ \emph {et~al.}(2017)\citenamefont {He},
  \citenamefont {Hartmann}, \citenamefont {Ma}, \citenamefont {Kim},
  \citenamefont {Ihly}, \citenamefont {Blackburn}, \citenamefont {Gao},
  \citenamefont {Kono}, \citenamefont {Yomogida}, \citenamefont {Hirano},
  \citenamefont {Tanaka}, \citenamefont {Kataura}, \citenamefont {Htoon},\ and\
  \citenamefont {Doorn}}]{He:2017}%
  \BibitemOpen
  \bibfield  {author} {\bibinfo {author} {\bibfnamefont {X.}~\bibnamefont
  {He}}, \bibinfo {author} {\bibfnamefont {N.~F.}\ \bibnamefont {Hartmann}},
  \bibinfo {author} {\bibfnamefont {X.}~\bibnamefont {Ma}}, \bibinfo {author}
  {\bibfnamefont {Y.}~\bibnamefont {Kim}}, \bibinfo {author} {\bibfnamefont
  {R.}~\bibnamefont {Ihly}}, \bibinfo {author} {\bibfnamefont {J.~L.}\
  \bibnamefont {Blackburn}}, \bibinfo {author} {\bibfnamefont {W.}~\bibnamefont
  {Gao}}, \bibinfo {author} {\bibfnamefont {J.}~\bibnamefont {Kono}}, \bibinfo
  {author} {\bibfnamefont {Y.}~\bibnamefont {Yomogida}}, \bibinfo {author}
  {\bibfnamefont {A.}~\bibnamefont {Hirano}}, \bibinfo {author} {\bibfnamefont
  {T.}~\bibnamefont {Tanaka}}, \bibinfo {author} {\bibfnamefont
  {H.}~\bibnamefont {Kataura}}, \bibinfo {author} {\bibfnamefont
  {H.}~\bibnamefont {Htoon}}, \ and\ \bibinfo {author} {\bibfnamefont {S.~K.}\
  \bibnamefont {Doorn}},\ }\bibfield  {title} {\bibinfo {title} {Tunable
  room-temperature single-photon emission at telecom wavelengths from $sp^3$
  defects in carbon nanotubes},\ }\href
  {http://dx.doi.org/10.1038/nphoton.2017.119} {\bibfield  {journal} {\bibinfo
  {journal} {Nat. Photon.}\ }\textbf {\bibinfo {volume} {11}},\ \bibinfo
  {pages} {577} (\bibinfo {year} {2017})}\BibitemShut {NoStop}%
\bibitem [{\citenamefont {Guldi}\ \emph {et~al.}(2005)\citenamefont {Guldi},
  \citenamefont {Rahman}, \citenamefont {Zerbetto},\ and\ \citenamefont
  {Prato}}]{Prato:2005}%
  \BibitemOpen
  \bibfield  {author} {\bibinfo {author} {\bibfnamefont {D.~M.}\ \bibnamefont
  {Guldi}}, \bibinfo {author} {\bibfnamefont {G.~M.~A.}\ \bibnamefont
  {Rahman}}, \bibinfo {author} {\bibfnamefont {F.}~\bibnamefont {Zerbetto}}, \
  and\ \bibinfo {author} {\bibfnamefont {M.}~\bibnamefont {Prato}},\ }\bibfield
   {title} {\bibinfo {title} {Carbon nanotubes in electron donor-acceptor
  nanocomposites},\ }\href {https://doi.org/10.1021/ar040238i} {\bibfield
  {journal} {\bibinfo  {journal} {Acc. Chem. Res.}\ }\textbf {\bibinfo {volume}
  {38}},\ \bibinfo {pages} {871} (\bibinfo {year} {2005})}\BibitemShut
  {NoStop}%
\bibitem [{\citenamefont {Ehli}\ \emph {et~al.}(2009)\citenamefont {Ehli},
  \citenamefont {Oelsner}, \citenamefont {Guldi}, \citenamefont {Mateo-Alonso},
  \citenamefont {Prato}, \citenamefont {Schmidt}, \citenamefont {Backes},
  \citenamefont {Hauke},\ and\ \citenamefont {Hirsch}}]{Hirsch:2009}%
  \BibitemOpen
  \bibfield  {author} {\bibinfo {author} {\bibfnamefont {C.}~\bibnamefont
  {Ehli}}, \bibinfo {author} {\bibfnamefont {C.}~\bibnamefont {Oelsner}},
  \bibinfo {author} {\bibfnamefont {D.~M.}\ \bibnamefont {Guldi}}, \bibinfo
  {author} {\bibfnamefont {A.}~\bibnamefont {Mateo-Alonso}}, \bibinfo {author}
  {\bibfnamefont {M.}~\bibnamefont {Prato}}, \bibinfo {author} {\bibfnamefont
  {C.}~\bibnamefont {Schmidt}}, \bibinfo {author} {\bibfnamefont
  {C.}~\bibnamefont {Backes}}, \bibinfo {author} {\bibfnamefont
  {F.}~\bibnamefont {Hauke}}, \ and\ \bibinfo {author} {\bibfnamefont
  {A.}~\bibnamefont {Hirsch}},\ }\bibfield  {title} {\bibinfo {title}
  {Manipulating single-wall carbon nanotubes by chemical doping and charge
  transfer with perylene dyes},\ }\href
  {https://www.nature.com/articles/nchem.214#supplementary-information}
  {\bibfield  {journal} {\bibinfo  {journal} {Nat. Chem.}\ }\textbf {\bibinfo
  {volume} {1}},\ \bibinfo {pages} {243} (\bibinfo {year} {2009})}\BibitemShut
  {NoStop}%
\bibitem [{\citenamefont {Malic}\ \emph {et~al.}(2011)\citenamefont {Malic},
  \citenamefont {Weber}, \citenamefont {Richter}, \citenamefont {Atalla},
  \citenamefont {Klamroth}, \citenamefont {Saalfrank}, \citenamefont {Reich},\
  and\ \citenamefont {Knorr}}]{Knorr:2011}%
  \BibitemOpen
  \bibfield  {author} {\bibinfo {author} {\bibfnamefont {E.}~\bibnamefont
  {Malic}}, \bibinfo {author} {\bibfnamefont {C.}~\bibnamefont {Weber}},
  \bibinfo {author} {\bibfnamefont {M.}~\bibnamefont {Richter}}, \bibinfo
  {author} {\bibfnamefont {V.}~\bibnamefont {Atalla}}, \bibinfo {author}
  {\bibfnamefont {T.}~\bibnamefont {Klamroth}}, \bibinfo {author}
  {\bibfnamefont {P.}~\bibnamefont {Saalfrank}}, \bibinfo {author}
  {\bibfnamefont {S.}~\bibnamefont {Reich}}, \ and\ \bibinfo {author}
  {\bibfnamefont {A.}~\bibnamefont {Knorr}},\ }\bibfield  {title} {\bibinfo
  {title} {Microscopic model of the optical absorption of carbon nanotubes
  functionalized with molecular spiropyran photoswitches},\ }\href {\doibase
  10.1103/PhysRevLett.106.097401} {\bibfield  {journal} {\bibinfo  {journal}
  {Phys. Rev. Lett.}\ }\textbf {\bibinfo {volume} {106}},\ \bibinfo {pages}
  {097401} (\bibinfo {year} {2011})}\BibitemShut {NoStop}%
\bibitem [{\citenamefont {Bartelmess}\ \emph {et~al.}(2010)\citenamefont
  {Bartelmess}, \citenamefont {Ballesteros}, \citenamefont {de~la Torre},
  \citenamefont {Kiessling}, \citenamefont {Campidelli}, \citenamefont {Prato},
  \citenamefont {Torres},\ and\ \citenamefont {Guldi}}]{Guldi:2010}%
  \BibitemOpen
  \bibfield  {author} {\bibinfo {author} {\bibfnamefont {J.}~\bibnamefont
  {Bartelmess}}, \bibinfo {author} {\bibfnamefont {B.}~\bibnamefont
  {Ballesteros}}, \bibinfo {author} {\bibfnamefont {G.}~\bibnamefont {de~la
  Torre}}, \bibinfo {author} {\bibfnamefont {D.}~\bibnamefont {Kiessling}},
  \bibinfo {author} {\bibfnamefont {S.}~\bibnamefont {Campidelli}}, \bibinfo
  {author} {\bibfnamefont {M.}~\bibnamefont {Prato}}, \bibinfo {author}
  {\bibfnamefont {T.}~\bibnamefont {Torres}}, \ and\ \bibinfo {author}
  {\bibfnamefont {D.~M.}\ \bibnamefont {Guldi}},\ }\bibfield  {title} {\bibinfo
  {title} {Phthalocyanine-pyrene conjugates: A powerful approach toward carbon
  nanotube solar cells},\ }\href {https://doi.org/10.1021/ja107131r} {\bibfield
   {journal} {\bibinfo  {journal} {J. Am. Chem. Soc.}\ }\textbf {\bibinfo
  {volume} {132}},\ \bibinfo {pages} {16202} (\bibinfo {year}
  {2010})}\BibitemShut {NoStop}%
\bibitem [{\citenamefont {Hecht}\ \emph {et~al.}(2006)\citenamefont {Hecht},
  \citenamefont {Ramirez}, \citenamefont {Briman}, \citenamefont {Artukovic},
  \citenamefont {Chichak}, \citenamefont {Stoddart},\ and\ \citenamefont
  {Gr\"uner}}]{Gruner:2006}%
  \BibitemOpen
  \bibfield  {author} {\bibinfo {author} {\bibfnamefont {D.~S.}\ \bibnamefont
  {Hecht}}, \bibinfo {author} {\bibfnamefont {R.~J.~A.}\ \bibnamefont
  {Ramirez}}, \bibinfo {author} {\bibfnamefont {M.}~\bibnamefont {Briman}},
  \bibinfo {author} {\bibfnamefont {E.}~\bibnamefont {Artukovic}}, \bibinfo
  {author} {\bibfnamefont {K.~S.}\ \bibnamefont {Chichak}}, \bibinfo {author}
  {\bibfnamefont {J.~F.}\ \bibnamefont {Stoddart}}, \ and\ \bibinfo {author}
  {\bibfnamefont {G.}~\bibnamefont {Gr\"uner}},\ }\bibfield  {title} {\bibinfo
  {title} {Bioinspired detection of light using a porphyrin-sensitized
  single-wall nanotube field effect transistor},\ }\href
  {https://doi.org/10.1021/nl061231s} {\bibfield  {journal} {\bibinfo
  {journal} {Nano Lett.}\ }\textbf {\bibinfo {volume} {6}},\ \bibinfo {pages}
  {2031} (\bibinfo {year} {2006})}\BibitemShut {NoStop}%
\bibitem [{\citenamefont {Alam}\ \emph {et~al.}(2017)\citenamefont {Alam},
  \citenamefont {Dehm}, \citenamefont {Hennrich}, \citenamefont {Zakharko},
  \citenamefont {Graf}, \citenamefont {Pfohl}, \citenamefont {Hossain},
  \citenamefont {Kappes}, \citenamefont {Zaumseil}, \citenamefont {Krupke},\
  and\ \citenamefont {Flavel}}]{Flavel:2017}%
  \BibitemOpen
  \bibfield  {author} {\bibinfo {author} {\bibfnamefont {A.}~\bibnamefont
  {Alam}}, \bibinfo {author} {\bibfnamefont {S.}~\bibnamefont {Dehm}}, \bibinfo
  {author} {\bibfnamefont {F.}~\bibnamefont {Hennrich}}, \bibinfo {author}
  {\bibfnamefont {Y.}~\bibnamefont {Zakharko}}, \bibinfo {author}
  {\bibfnamefont {A.}~\bibnamefont {Graf}}, \bibinfo {author} {\bibfnamefont
  {M.}~\bibnamefont {Pfohl}}, \bibinfo {author} {\bibfnamefont {I.~M.}\
  \bibnamefont {Hossain}}, \bibinfo {author} {\bibfnamefont {M.~M.}\
  \bibnamefont {Kappes}}, \bibinfo {author} {\bibfnamefont {J.}~\bibnamefont
  {Zaumseil}}, \bibinfo {author} {\bibfnamefont {R.}~\bibnamefont {Krupke}}, \
  and\ \bibinfo {author} {\bibfnamefont {B.~S.}\ \bibnamefont {Flavel}},\
  }\bibfield  {title} {\bibinfo {title} {Photocurrent spectroscopy of
  dye-sensitized carbon nanotubes},\ }\href {\doibase 10.1039/C7NR04022A}
  {\bibfield  {journal} {\bibinfo  {journal} {Nanoscale}\ }\textbf {\bibinfo
  {volume} {9}},\ \bibinfo {pages} {11205} (\bibinfo {year}
  {2017})}\BibitemShut {NoStop}%
\bibitem [{\citenamefont {Roquelet}\ \emph {et~al.}(2010)\citenamefont
  {Roquelet}, \citenamefont {Garrot}, \citenamefont {Lauret}, \citenamefont
  {Voisin}, \citenamefont {Alain-Rizzo}, \citenamefont {Roussignol},
  \citenamefont {Delaire},\ and\ \citenamefont {Deleporte}}]{Deleporte:2010}%
  \BibitemOpen
  \bibfield  {author} {\bibinfo {author} {\bibfnamefont {C.}~\bibnamefont
  {Roquelet}}, \bibinfo {author} {\bibfnamefont {D.}~\bibnamefont {Garrot}},
  \bibinfo {author} {\bibfnamefont {J.~S.}\ \bibnamefont {Lauret}}, \bibinfo
  {author} {\bibfnamefont {C.}~\bibnamefont {Voisin}}, \bibinfo {author}
  {\bibfnamefont {V.}~\bibnamefont {Alain-Rizzo}}, \bibinfo {author}
  {\bibfnamefont {P.}~\bibnamefont {Roussignol}}, \bibinfo {author}
  {\bibfnamefont {J.~A.}\ \bibnamefont {Delaire}}, \ and\ \bibinfo {author}
  {\bibfnamefont {E.}~\bibnamefont {Deleporte}},\ }\bibfield  {title} {\bibinfo
  {title} {Quantum efficiency of energy transfer in noncovalent carbon
  nanotube/porphyrin compounds},\ }\href {https://doi.org/10.1063/1.3496470}
  {\bibfield  {journal} {\bibinfo  {journal} {Appl. Phys. Lett.}\ }\textbf
  {\bibinfo {volume} {97}},\ \bibinfo {pages} {141918} (\bibinfo {year}
  {2010})}\BibitemShut {NoStop}%
\bibitem [{\citenamefont {Correa}\ and\ \citenamefont
  {Orellana}(2012)}]{Orellana:2012}%
  \BibitemOpen
  \bibfield  {author} {\bibinfo {author} {\bibfnamefont {J.~D.}\ \bibnamefont
  {Correa}}\ and\ \bibinfo {author} {\bibfnamefont {W.}~\bibnamefont
  {Orellana}},\ }\bibfield  {title} {\bibinfo {title} {Optical response of
  carbon nanotubes functionalized with (free-base, zn) porphyrins, and
  phthalocyanines: A dft study},\ }\href {\doibase 10.1103/PhysRevB.86.125417}
  {\bibfield  {journal} {\bibinfo  {journal} {Phys. Rev. B}\ }\textbf {\bibinfo
  {volume} {86}},\ \bibinfo {pages} {125417} (\bibinfo {year}
  {2012})}\BibitemShut {NoStop}%
\bibitem [{\citenamefont {Orellana}(2014)}]{Orellana:2014}%
  \BibitemOpen
  \bibfield  {author} {\bibinfo {author} {\bibfnamefont {W.}~\bibnamefont
  {Orellana}},\ }\bibfield  {title} {\bibinfo {title} {Strong $\pi$-$\pi$
  interaction of porphyrins on (6,5) carbon nanotubes with full surface
  coverage: Ab-initio calculations},\ }\href
  {https://doi.org/10.1063/1.4890591} {\bibfield  {journal} {\bibinfo
  {journal} {Appl. Phys. Lett.}\ }\textbf {\bibinfo {volume} {105}},\ \bibinfo
  {pages} {023110} (\bibinfo {year} {2014})}\BibitemShut {NoStop}%
\bibitem [{\citenamefont {Garrot}\ \emph {et~al.}(2011)\citenamefont {Garrot},
  \citenamefont {Langlois}, \citenamefont {Roquelet}, \citenamefont {Michel},
  \citenamefont {Roussignol}, \citenamefont {Delalande}, \citenamefont
  {Deleporte}, \citenamefont {Lauret},\ and\ \citenamefont
  {Voisin}}]{Voisin:2011}%
  \BibitemOpen
  \bibfield  {author} {\bibinfo {author} {\bibfnamefont {D.}~\bibnamefont
  {Garrot}}, \bibinfo {author} {\bibfnamefont {B.}~\bibnamefont {Langlois}},
  \bibinfo {author} {\bibfnamefont {C.}~\bibnamefont {Roquelet}}, \bibinfo
  {author} {\bibfnamefont {T.}~\bibnamefont {Michel}}, \bibinfo {author}
  {\bibfnamefont {P.}~\bibnamefont {Roussignol}}, \bibinfo {author}
  {\bibfnamefont {C.}~\bibnamefont {Delalande}}, \bibinfo {author}
  {\bibfnamefont {E.}~\bibnamefont {Deleporte}}, \bibinfo {author}
  {\bibfnamefont {J.~S.}\ \bibnamefont {Lauret}}, \ and\ \bibinfo {author}
  {\bibfnamefont {C.}~\bibnamefont {Voisin}},\ }\bibfield  {title} {\bibinfo
  {title} {Time-resolved investigation of excitation energy transfer in carbon
  nanotube-porphyrin compounds},\ }\href {https://doi.org/10.1021/jp207267e}
  {\bibfield  {journal} {\bibinfo  {journal} {J. Phys. Chem. C}\ }\textbf
  {\bibinfo {volume} {115}},\ \bibinfo {pages} {23283} (\bibinfo {year}
  {2011})}\BibitemShut {NoStop}%
\bibitem [{\citenamefont {Casey}\ \emph {et~al.}(2008)\citenamefont {Casey},
  \citenamefont {Bachilo},\ and\ \citenamefont {Weisman}}]{Bruce:2008}%
  \BibitemOpen
  \bibfield  {author} {\bibinfo {author} {\bibfnamefont {J.~P.}\ \bibnamefont
  {Casey}}, \bibinfo {author} {\bibfnamefont {S.~M.}\ \bibnamefont {Bachilo}},
  \ and\ \bibinfo {author} {\bibfnamefont {R.~B.}\ \bibnamefont {Weisman}},\
  }\bibfield  {title} {\bibinfo {title} {Efficient photosensitized energy
  transfer and near-ir fluorescence from porphyrin-swnt complexes},\ }\href
  {\doibase 10.1039/B716649D} {\bibfield  {journal} {\bibinfo  {journal} {J.
  Mater. Chem.}\ }\textbf {\bibinfo {volume} {18}},\ \bibinfo {pages} {1510}
  (\bibinfo {year} {2008})}\BibitemShut {NoStop}%
\bibitem [{\citenamefont {Zhong}\ \emph {et~al.}(2013)\citenamefont {Zhong},
  \citenamefont {Diev}, \citenamefont {Roberts}, \citenamefont {Antunez},
  \citenamefont {Brutchey}, \citenamefont {Bradforth},\ and\ \citenamefont
  {Thompson}}]{Thompson:2013}%
  \BibitemOpen
  \bibfield  {author} {\bibinfo {author} {\bibfnamefont {Q.}~\bibnamefont
  {Zhong}}, \bibinfo {author} {\bibfnamefont {V.~V.}\ \bibnamefont {Diev}},
  \bibinfo {author} {\bibfnamefont {S.~T.}\ \bibnamefont {Roberts}}, \bibinfo
  {author} {\bibfnamefont {P.~D.}\ \bibnamefont {Antunez}}, \bibinfo {author}
  {\bibfnamefont {R.~L.}\ \bibnamefont {Brutchey}}, \bibinfo {author}
  {\bibfnamefont {S.~E.}\ \bibnamefont {Bradforth}}, \ and\ \bibinfo {author}
  {\bibfnamefont {M.~E.}\ \bibnamefont {Thompson}},\ }\bibfield  {title}
  {\bibinfo {title} {Fused porphyrin-single-walled carbon nanotube hybrids:
  Efficient formation and photophysical characterization},\ }\href
  {https://doi.org/10.1021/nn400362e} {\bibfield  {journal} {\bibinfo
  {journal} {ACS Nano}\ }\textbf {\bibinfo {volume} {7}},\ \bibinfo {pages}
  {3466} (\bibinfo {year} {2013})}\BibitemShut {NoStop}%
\bibitem [{\citenamefont {Sarkar}\ \emph {et~al.}(2018)\citenamefont {Sarkar},
  \citenamefont {Habib}, \citenamefont {Pal},\ and\ \citenamefont
  {Prezhdo}}]{Prezhdo:2018}%
  \BibitemOpen
  \bibfield  {author} {\bibinfo {author} {\bibfnamefont {R.}~\bibnamefont
  {Sarkar}}, \bibinfo {author} {\bibfnamefont {M.}~\bibnamefont {Habib}},
  \bibinfo {author} {\bibfnamefont {S.}~\bibnamefont {Pal}}, \ and\ \bibinfo
  {author} {\bibfnamefont {O.~V.}\ \bibnamefont {Prezhdo}},\ }\bibfield
  {title} {\bibinfo {title} {Ultrafast, asymmetric charge transfer and slow
  charge recombination in porphyrin/cnt composites demonstrated by time-domain
  atomistic simulation},\ }\href {http://dx.doi.org/10.1039/C8NR02544D}
  {\bibfield  {journal} {\bibinfo  {journal} {Nanoscale}\ }\textbf {\bibinfo
  {volume} {10}},\ \bibinfo {pages} {12683} (\bibinfo {year}
  {2018})}\BibitemShut {NoStop}%
\bibitem [{\citenamefont {Stranks}\ \emph {et~al.}(2011)\citenamefont
  {Stranks}, \citenamefont {Sprafke}, \citenamefont {Anderson},\ and\
  \citenamefont {Nicholas}}]{Nichola:2011}%
  \BibitemOpen
  \bibfield  {author} {\bibinfo {author} {\bibfnamefont {S.~D.}\ \bibnamefont
  {Stranks}}, \bibinfo {author} {\bibfnamefont {J.~K.}\ \bibnamefont
  {Sprafke}}, \bibinfo {author} {\bibfnamefont {H.~L.}\ \bibnamefont
  {Anderson}}, \ and\ \bibinfo {author} {\bibfnamefont {R.~J.}\ \bibnamefont
  {Nicholas}},\ }\bibfield  {title} {\bibinfo {title} {Electronic and
  mechanical modification of single-walled carbon nanotubes by binding to
  porphyrin oligomers},\ }\href {\doibase 10.1021/nn103588h} {\bibfield
  {journal} {\bibinfo  {journal} {ACS Nano}\ }\textbf {\bibinfo {volume} {5}},\
  \bibinfo {pages} {2307} (\bibinfo {year} {2011})}\BibitemShut {NoStop}%
\bibitem [{\citenamefont {Lozzi}\ \emph {et~al.}(2008)\citenamefont {Lozzi},
  \citenamefont {Santucci}, \citenamefont {Bussolotti},\ and\ \citenamefont
  {La~Rosa}}]{Rosa:2008}%
  \BibitemOpen
  \bibfield  {author} {\bibinfo {author} {\bibfnamefont {L.}~\bibnamefont
  {Lozzi}}, \bibinfo {author} {\bibfnamefont {S.}~\bibnamefont {Santucci}},
  \bibinfo {author} {\bibfnamefont {F.}~\bibnamefont {Bussolotti}}, \ and\
  \bibinfo {author} {\bibfnamefont {S.}~\bibnamefont {La~Rosa}},\ }\bibfield
  {title} {\bibinfo {title} {Investigation on copper phthalocyanine/multiwalled
  carbon nanotube interface},\ }\href {https://doi.org/10.1063/1.2961325}
  {\bibfield  {journal} {\bibinfo  {journal} {J. Appl. Phys.}\ }\textbf
  {\bibinfo {volume} {104}},\ \bibinfo {pages} {033701} (\bibinfo {year}
  {2008})}\BibitemShut {NoStop}%
\bibitem [{\citenamefont {Basiuk}\ \emph {et~al.}(2018)\citenamefont {Basiuk},
  \citenamefont {Flores-Sánchez}, \citenamefont {Meza-Laguna}, \citenamefont
  {Flores-Flores}, \citenamefont {Bucio-Galindo}, \citenamefont {Puente-Lee},\
  and\ \citenamefont {Basiuk}}]{Basiuk:2018}%
  \BibitemOpen
  \bibfield  {author} {\bibinfo {author} {\bibfnamefont {V.~A.}\ \bibnamefont
  {Basiuk}}, \bibinfo {author} {\bibfnamefont {L.~J.}\ \bibnamefont
  {Flores-Sánchez}}, \bibinfo {author} {\bibfnamefont {V.}~\bibnamefont
  {Meza-Laguna}}, \bibinfo {author} {\bibfnamefont {J.~O.}\ \bibnamefont
  {Flores-Flores}}, \bibinfo {author} {\bibfnamefont {L.}~\bibnamefont
  {Bucio-Galindo}}, \bibinfo {author} {\bibfnamefont {I.}~\bibnamefont
  {Puente-Lee}}, \ and\ \bibinfo {author} {\bibfnamefont {E.~V.}\ \bibnamefont
  {Basiuk}},\ }\bibfield  {title} {\bibinfo {title} {Noncovalent
  functionalization of pristine cvd single-walled carbon nanotubes with 3d
  metal(ii) phthalocyanines by adsorption from the gas phase},\ }\href
  {\doibase https://doi.org/10.1016/j.apsusc.2017.12.122} {\bibfield  {journal}
  {\bibinfo  {journal} {Appl. Surf. Sci.}\ }\textbf {\bibinfo {volume} {436}},\
  \bibinfo {pages} {1123 } (\bibinfo {year} {2018})}\BibitemShut {NoStop}%
\bibitem [{\citenamefont {Ishii}\ \emph {et~al.}(2015)\citenamefont {Ishii},
  \citenamefont {Yoshida},\ and\ \citenamefont {Kato}}]{Ishii:2015}%
  \BibitemOpen
  \bibfield  {author} {\bibinfo {author} {\bibfnamefont {A.}~\bibnamefont
  {Ishii}}, \bibinfo {author} {\bibfnamefont {M.}~\bibnamefont {Yoshida}}, \
  and\ \bibinfo {author} {\bibfnamefont {Y.~K.}\ \bibnamefont {Kato}},\
  }\bibfield  {title} {\bibinfo {title} {Exciton diffusion, end quenching, and
  exciton-exciton annihilation in individual air-suspended carbon nanotubes},\
  }\href {\doibase 10.1103/PhysRevB.91.125427} {\bibfield  {journal} {\bibinfo
  {journal} {Phys. Rev. B}\ }\textbf {\bibinfo {volume} {91}},\ \bibinfo
  {pages} {125427} (\bibinfo {year} {2015})}\BibitemShut {NoStop}%
\bibitem [{\citenamefont {Tang}(1986)}]{Tang:1986}%
  \BibitemOpen
  \bibfield  {author} {\bibinfo {author} {\bibfnamefont {C.~W.}\ \bibnamefont
  {Tang}},\ }\bibfield  {title} {\bibinfo {title} {Two-layer organic
  photovoltaic cell},\ }\href {https://doi.org/10.1063/1.96937} {\bibfield
  {journal} {\bibinfo  {journal} {Appl. Phys. Lett.}\ }\textbf {\bibinfo
  {volume} {48}},\ \bibinfo {pages} {183} (\bibinfo {year} {1986})}\BibitemShut
  {NoStop}%
\bibitem [{\citenamefont {Uda}\ \emph {et~al.}(2018)\citenamefont {Uda},
  \citenamefont {Tanaka},\ and\ \citenamefont {Kato}}]{Uda:2018a}%
  \BibitemOpen
  \bibfield  {author} {\bibinfo {author} {\bibfnamefont {T.}~\bibnamefont
  {Uda}}, \bibinfo {author} {\bibfnamefont {S.}~\bibnamefont {Tanaka}}, \ and\
  \bibinfo {author} {\bibfnamefont {Y.~K.}\ \bibnamefont {Kato}},\ }\bibfield
  {title} {\bibinfo {title} {Molecular screening effects on exciton-carrier
  interactions in suspended carbon nanotubes},\ }\href
  {https://doi.org/10.1063/1.5046433} {\bibfield  {journal} {\bibinfo
  {journal} {Appl. Phys. Lett.}\ }\textbf {\bibinfo {volume} {113}},\ \bibinfo
  {pages} {121105} (\bibinfo {year} {2018})}\BibitemShut {NoStop}%
\bibitem [{\citenamefont {Lefebvre}\ and\ \citenamefont
  {Finnie}(2008)}]{Lefebvre:2008}%
  \BibitemOpen
  \bibfield  {author} {\bibinfo {author} {\bibfnamefont {J.}~\bibnamefont
  {Lefebvre}}\ and\ \bibinfo {author} {\bibfnamefont {P.}~\bibnamefont
  {Finnie}},\ }\bibfield  {title} {\bibinfo {title} {Excited excitonic states
  in single-walled carbon nanotubes},\ }\href {\doibase 10.1021/nl080518h}
  {\bibfield  {journal} {\bibinfo  {journal} {Nano Lett.}\ }\textbf {\bibinfo
  {volume} {8}},\ \bibinfo {pages} {1890} (\bibinfo {year} {2008})}\BibitemShut
  {NoStop}%
\bibitem [{\citenamefont {Larsen}\ \emph {et~al.}(2012)\citenamefont {Larsen},
  \citenamefont {Deria}, \citenamefont {Holt}, \citenamefont {Stanton},
  \citenamefont {Heben}, \citenamefont {Therien},\ and\ \citenamefont
  {Blackburn}}]{Blackburn:2012}%
  \BibitemOpen
  \bibfield  {author} {\bibinfo {author} {\bibfnamefont {B.~A.}\ \bibnamefont
  {Larsen}}, \bibinfo {author} {\bibfnamefont {P.}~\bibnamefont {Deria}},
  \bibinfo {author} {\bibfnamefont {J.~M.}\ \bibnamefont {Holt}}, \bibinfo
  {author} {\bibfnamefont {I.~N.}\ \bibnamefont {Stanton}}, \bibinfo {author}
  {\bibfnamefont {M.~J.}\ \bibnamefont {Heben}}, \bibinfo {author}
  {\bibfnamefont {M.~J.}\ \bibnamefont {Therien}}, \ and\ \bibinfo {author}
  {\bibfnamefont {J.~L.}\ \bibnamefont {Blackburn}},\ }\bibfield  {title}
  {\bibinfo {title} {Effect of solvent polarity and electrophilicity on quantum
  yields and solvatochromic shifts of single-walled carbon nanotube
  photoluminescence},\ }\href {https://doi.org/10.1021/ja2114618} {\bibfield
  {journal} {\bibinfo  {journal} {J. Am. Chem. Soc.}\ }\textbf {\bibinfo
  {volume} {134}},\ \bibinfo {pages} {12485} (\bibinfo {year}
  {2012})}\BibitemShut {NoStop}%
\bibitem [{\citenamefont {Yasukochi}\ \emph {et~al.}(2011)\citenamefont
  {Yasukochi}, \citenamefont {Murai}, \citenamefont {Moritsubo}, \citenamefont
  {Shimada}, \citenamefont {Chiashi}, \citenamefont {Maruyama},\ and\
  \citenamefont {Kato}}]{Yasukochi:2011}%
  \BibitemOpen
  \bibfield  {author} {\bibinfo {author} {\bibfnamefont {S.}~\bibnamefont
  {Yasukochi}}, \bibinfo {author} {\bibfnamefont {T.}~\bibnamefont {Murai}},
  \bibinfo {author} {\bibfnamefont {S.}~\bibnamefont {Moritsubo}}, \bibinfo
  {author} {\bibfnamefont {T.}~\bibnamefont {Shimada}}, \bibinfo {author}
  {\bibfnamefont {S.}~\bibnamefont {Chiashi}}, \bibinfo {author} {\bibfnamefont
  {S.}~\bibnamefont {Maruyama}}, \ and\ \bibinfo {author} {\bibfnamefont
  {Y.~K.}\ \bibnamefont {Kato}},\ }\bibfield  {title} {\bibinfo {title}
  {Gate-induced blueshift and quenching of photoluminescence in suspended
  single-walled carbon nanotubes},\ }\href {\doibase
  10.1103/PhysRevB.84.121409} {\bibfield  {journal} {\bibinfo  {journal} {Phys.
  Rev. B}\ }\textbf {\bibinfo {volume} {84}},\ \bibinfo {pages} {121409(R)}
  (\bibinfo {year} {2011})}\BibitemShut {NoStop}%
\bibitem [{\citenamefont {Yoshida}\ \emph {et~al.}(2016)\citenamefont
  {Yoshida}, \citenamefont {Popert},\ and\ \citenamefont
  {Kato}}]{Yoshida:2016}%
  \BibitemOpen
  \bibfield  {author} {\bibinfo {author} {\bibfnamefont {M.}~\bibnamefont
  {Yoshida}}, \bibinfo {author} {\bibfnamefont {A.}~\bibnamefont {Popert}}, \
  and\ \bibinfo {author} {\bibfnamefont {Y.~K.}\ \bibnamefont {Kato}},\
  }\bibfield  {title} {\bibinfo {title} {Gate-voltage induced trions in
  suspended carbon nanotubes},\ }\href {\doibase 10.1103/PhysRevB.93.041402}
  {\bibfield  {journal} {\bibinfo  {journal} {Phys. Rev. B}\ }\textbf {\bibinfo
  {volume} {93}},\ \bibinfo {pages} {041402(R)} (\bibinfo {year}
  {2016})}\BibitemShut {NoStop}%
\bibitem [{\citenamefont {Schuettfort}\ \emph {et~al.}(2009)\citenamefont
  {Schuettfort}, \citenamefont {Nish},\ and\ \citenamefont
  {Nicholas}}]{Nicholas:2009}%
  \BibitemOpen
  \bibfield  {author} {\bibinfo {author} {\bibfnamefont {T.}~\bibnamefont
  {Schuettfort}}, \bibinfo {author} {\bibfnamefont {A.}~\bibnamefont {Nish}}, \
  and\ \bibinfo {author} {\bibfnamefont {R.~J.}\ \bibnamefont {Nicholas}},\
  }\bibfield  {title} {\bibinfo {title} {Observation of a type ii
  heterojunction in a highly ordered polymer-carbon nanotube nanohybrid
  structure},\ }\href {https://doi.org/10.1021/nl902081t} {\bibfield  {journal}
  {\bibinfo  {journal} {Nano Lett.}\ }\textbf {\bibinfo {volume} {9}},\
  \bibinfo {pages} {3871} (\bibinfo {year} {2009})}\BibitemShut {NoStop}%
\bibitem [{\citenamefont {Choi}\ and\ \citenamefont
  {Strano}(2007)}]{Strano:2007}%
  \BibitemOpen
  \bibfield  {author} {\bibinfo {author} {\bibfnamefont {J.~H.}\ \bibnamefont
  {Choi}}\ and\ \bibinfo {author} {\bibfnamefont {M.~S.}\ \bibnamefont
  {Strano}},\ }\bibfield  {title} {\bibinfo {title} {Solvatochromism in
  single-walled carbon nanotubes},\ }\href {https://doi.org/10.1063/1.2745228}
  {\bibfield  {journal} {\bibinfo  {journal} {Appl. Phys. Lett.}\ }\textbf
  {\bibinfo {volume} {90}},\ \bibinfo {pages} {223114} (\bibinfo {year}
  {2007})}\BibitemShut {NoStop}%
\bibitem [{\citenamefont {Ohno}\ \emph {et~al.}(2007)\citenamefont {Ohno},
  \citenamefont {Iwasaki}, \citenamefont {Murakami}, \citenamefont {Kishimoto},
  \citenamefont {Maruyama},\ and\ \citenamefont {Mizutani}}]{Ohno:2007}%
  \BibitemOpen
  \bibfield  {author} {\bibinfo {author} {\bibfnamefont {Y.}~\bibnamefont
  {Ohno}}, \bibinfo {author} {\bibfnamefont {S.}~\bibnamefont {Iwasaki}},
  \bibinfo {author} {\bibfnamefont {Y.}~\bibnamefont {Murakami}}, \bibinfo
  {author} {\bibfnamefont {S.}~\bibnamefont {Kishimoto}}, \bibinfo {author}
  {\bibfnamefont {S.}~\bibnamefont {Maruyama}}, \ and\ \bibinfo {author}
  {\bibfnamefont {T.}~\bibnamefont {Mizutani}},\ }\bibfield  {title} {\bibinfo
  {title} {Excitonic transition energies in single-walled carbon nanotubes:
  Dependence on environmental dielectric constant},\ }\href {\doibase
  10.1002/pssb.200776124} {\bibfield  {journal} {\bibinfo  {journal} {Phys.
  Status Solidi B}\ }\textbf {\bibinfo {volume} {244}},\ \bibinfo {pages}
  {4002} (\bibinfo {year} {2007})}\BibitemShut {NoStop}%
\bibitem [{\citenamefont {Walsh}\ \emph {et~al.}(2007)\citenamefont {Walsh},
  \citenamefont {Vamivakas}, \citenamefont {Yin}, \citenamefont {Cronin},
  \citenamefont {\"Unl\"u}, \citenamefont {Goldberg},\ and\ \citenamefont
  {Swan}}]{Swan:2007}%
  \BibitemOpen
  \bibfield  {author} {\bibinfo {author} {\bibfnamefont {A.~G.}\ \bibnamefont
  {Walsh}}, \bibinfo {author} {\bibfnamefont {A.~N.}\ \bibnamefont
  {Vamivakas}}, \bibinfo {author} {\bibfnamefont {Y.}~\bibnamefont {Yin}},
  \bibinfo {author} {\bibfnamefont {S.~B.}\ \bibnamefont {Cronin}}, \bibinfo
  {author} {\bibfnamefont {M.~S.}\ \bibnamefont {\"Unl\"u}}, \bibinfo {author}
  {\bibfnamefont {B.~B.}\ \bibnamefont {Goldberg}}, \ and\ \bibinfo {author}
  {\bibfnamefont {A.~K.}\ \bibnamefont {Swan}},\ }\bibfield  {title} {\bibinfo
  {title} {Screening of excitons in single, suspended carbon nanotubes},\
  }\href {https://doi.org/10.1021/nl070193p} {\bibfield  {journal} {\bibinfo
  {journal} {Nano Lett.}\ }\textbf {\bibinfo {volume} {7}},\ \bibinfo {pages}
  {1485} (\bibinfo {year} {2007})}\BibitemShut {NoStop}%
\bibitem [{\citenamefont {Miyauchi}\ \emph {et~al.}(2015)\citenamefont
  {Miyauchi}, \citenamefont {Zhang}, \citenamefont {Takekoshi}, \citenamefont
  {Tomio}, \citenamefont {Suzuura}, \citenamefont {Perebeinos}, \citenamefont
  {Deshpande}, \citenamefont {Lu}, \citenamefont {Berciaud}, \citenamefont
  {Kim}, \citenamefont {Hone},\ and\ \citenamefont {Heinz}}]{Miyauchi:2015}%
  \BibitemOpen
  \bibfield  {author} {\bibinfo {author} {\bibfnamefont {Y.}~\bibnamefont
  {Miyauchi}}, \bibinfo {author} {\bibfnamefont {Z.}~\bibnamefont {Zhang}},
  \bibinfo {author} {\bibfnamefont {M.}~\bibnamefont {Takekoshi}}, \bibinfo
  {author} {\bibfnamefont {Y.}~\bibnamefont {Tomio}}, \bibinfo {author}
  {\bibfnamefont {H.}~\bibnamefont {Suzuura}}, \bibinfo {author} {\bibfnamefont
  {V.}~\bibnamefont {Perebeinos}}, \bibinfo {author} {\bibfnamefont {V.~V.}\
  \bibnamefont {Deshpande}}, \bibinfo {author} {\bibfnamefont {C.}~\bibnamefont
  {Lu}}, \bibinfo {author} {\bibfnamefont {S.}~\bibnamefont {Berciaud}},
  \bibinfo {author} {\bibfnamefont {P.}~\bibnamefont {Kim}}, \bibinfo {author}
  {\bibfnamefont {J.}~\bibnamefont {Hone}}, \ and\ \bibinfo {author}
  {\bibfnamefont {T.~F.}\ \bibnamefont {Heinz}},\ }\bibfield  {title} {\bibinfo
  {title} {Tunable electronic correlation effects in nanotube-light
  interactions},\ }\href {\doibase 10.1103/PhysRevB.92.205407} {\bibfield
  {journal} {\bibinfo  {journal} {Phys. Rev. B}\ }\textbf {\bibinfo {volume}
  {92}},\ \bibinfo {pages} {205407} (\bibinfo {year} {2015})}\BibitemShut
  {NoStop}%
\bibitem [{\citenamefont {Sato}\ \emph {et~al.}(2007)\citenamefont {Sato},
  \citenamefont {Saito}, \citenamefont {Jiang}, \citenamefont {Dresselhaus},\
  and\ \citenamefont {Dresselhaus}}]{Dresselhaus:2007}%
  \BibitemOpen
  \bibfield  {author} {\bibinfo {author} {\bibfnamefont {K.}~\bibnamefont
  {Sato}}, \bibinfo {author} {\bibfnamefont {R.}~\bibnamefont {Saito}},
  \bibinfo {author} {\bibfnamefont {J.}~\bibnamefont {Jiang}}, \bibinfo
  {author} {\bibfnamefont {G.}~\bibnamefont {Dresselhaus}}, \ and\ \bibinfo
  {author} {\bibfnamefont {M.~S.}\ \bibnamefont {Dresselhaus}},\ }\bibfield
  {title} {\bibinfo {title} {Discontinuity in the family pattern of single-wall
  carbon nanotubes},\ }\href {\doibase 10.1103/PhysRevB.76.195446} {\bibfield
  {journal} {\bibinfo  {journal} {Phys. Rev. B}\ }\textbf {\bibinfo {volume}
  {76}},\ \bibinfo {pages} {195446} (\bibinfo {year} {2007})}\BibitemShut
  {NoStop}%
\bibitem [{\citenamefont {Matsunaga}\ \emph {et~al.}(2010)\citenamefont
  {Matsunaga}, \citenamefont {Matsuda},\ and\ \citenamefont
  {Kanemitsu}}]{Matsunaga:2010}%
  \BibitemOpen
  \bibfield  {author} {\bibinfo {author} {\bibfnamefont {R.}~\bibnamefont
  {Matsunaga}}, \bibinfo {author} {\bibfnamefont {K.}~\bibnamefont {Matsuda}},
  \ and\ \bibinfo {author} {\bibfnamefont {Y.}~\bibnamefont {Kanemitsu}},\
  }\bibfield  {title} {\bibinfo {title} {Origin of low-energy photoluminescence
  peaks in single carbon nanotubes: ${K}$-momentum dark excitons and triplet
  dark excitons},\ }\href {\doibase 10.1103/PhysRevB.81.033401} {\bibfield
  {journal} {\bibinfo  {journal} {Phys. Rev. B}\ }\textbf {\bibinfo {volume}
  {81}},\ \bibinfo {pages} {033401} (\bibinfo {year} {2010})}\BibitemShut
  {NoStop}%
\bibitem [{\citenamefont {Matsunaga}\ \emph {et~al.}(2011)\citenamefont
  {Matsunaga}, \citenamefont {Matsuda},\ and\ \citenamefont
  {Kanemitsu}}]{Matsunaga:2011}%
  \BibitemOpen
  \bibfield  {author} {\bibinfo {author} {\bibfnamefont {R.}~\bibnamefont
  {Matsunaga}}, \bibinfo {author} {\bibfnamefont {K.}~\bibnamefont {Matsuda}},
  \ and\ \bibinfo {author} {\bibfnamefont {Y.}~\bibnamefont {Kanemitsu}},\
  }\bibfield  {title} {\bibinfo {title} {Observation of charged excitons in
  hole-doped carbon nanotubes using photoluminescence and absorption
  spectroscopy},\ }\href {\doibase 10.1103/PhysRevLett.106.037404} {\bibfield
  {journal} {\bibinfo  {journal} {Phys. Rev. Lett.}\ }\textbf {\bibinfo
  {volume} {106}},\ \bibinfo {pages} {037404} (\bibinfo {year}
  {2011})}\BibitemShut {NoStop}%
\bibitem [{\citenamefont {Santos}\ \emph {et~al.}(2011)\citenamefont {Santos},
  \citenamefont {Yuma}, \citenamefont {Berciaud}, \citenamefont {Shaver},
  \citenamefont {Gallart}, \citenamefont {Gilliot}, \citenamefont {Cognet},\
  and\ \citenamefont {Lounis}}]{Santos:2011}%
  \BibitemOpen
  \bibfield  {author} {\bibinfo {author} {\bibfnamefont {S.~M.}\ \bibnamefont
  {Santos}}, \bibinfo {author} {\bibfnamefont {B.}~\bibnamefont {Yuma}},
  \bibinfo {author} {\bibfnamefont {S.}~\bibnamefont {Berciaud}}, \bibinfo
  {author} {\bibfnamefont {J.}~\bibnamefont {Shaver}}, \bibinfo {author}
  {\bibfnamefont {M.}~\bibnamefont {Gallart}}, \bibinfo {author} {\bibfnamefont
  {P.}~\bibnamefont {Gilliot}}, \bibinfo {author} {\bibfnamefont
  {L.}~\bibnamefont {Cognet}}, \ and\ \bibinfo {author} {\bibfnamefont
  {B.}~\bibnamefont {Lounis}},\ }\bibfield  {title} {\bibinfo {title}
  {All-optical trion generation in single-walled carbon nanotubes},\ }\href
  {\doibase 10.1103/PhysRevLett.107.187401} {\bibfield  {journal} {\bibinfo
  {journal} {Phys. Rev. Lett.}\ }\textbf {\bibinfo {volume} {107}},\ \bibinfo
  {pages} {187401} (\bibinfo {year} {2011})}\BibitemShut {NoStop}%
\end{thebibliography}
\end{document}